\documentclass[journal,twoside]{IEEEtran}
\usepackage{graphicx}           
\usepackage{amsmath,amsfonts}
\usepackage{url}
\usepackage{times,epsfig}
\usepackage{psfrag}
\usepackage{subfigure}
\usepackage{stfloats}
\usepackage{amssymb}
\usepackage{multirow}
\usepackage[justification=centering]{caption}
\usepackage{multirow}
\usepackage{cite}
\usepackage{CJK}
\usepackage{url}
\usepackage{algorithmic}
\usepackage{tabu}
\usepackage[lined,ruled]{algorithm2e}
\usepackage{caption}
\usepackage{float}

\ifCLASSINFOpdf
\else
\fi
\begin{document}
\bibliographystyle{ieeetr}
\title{Permissioned Blockchain for Efficient and Secure Resource Sharing in Vehicular Edge Computing}

\author{Siming Wang, Xumin Huang, Rong Yu, \emph{Member, IEEE}, Yan Zhang, \emph{Senior Member, IEEE}, \\ and  Ekram Hossain,  \emph{Fellow, IEEE}
\IEEEcompsocitemizethanks{

\IEEEcompsocthanksitem Siming Wang, Xumin Huang and Rong Yu are with School of Automation, Guangdong University of Technology, China. 
Email: simingwang30@163.com; huangxu\_min@163.com; yurong@ieee.org.
\IEEEcompsocthanksitem Yan Zhang is with the University of Oslo, Norway, and also with Simula Research Laboratory, Norway (e-mail: yanzhang@ieee.org).
\IEEEcompsocthanksitem Ekram Hossain is with the Department of Electrical and Computer Engineering, University of Manitoba, Winnipeg, MB R3T 2N2, Canada
(e-mail: Ekram.Hossain@umanitoba.ca). }
}
\maketitle
\thispagestyle{empty}
\begin{abstract}
With the fast expanding scale of vehicular networks, vehicular edge computing (VEC) has emerged and attracted growing attention from both industry and academia. Parked vehicles (PVs) have great potential to join vehicular networks for sharing their idle computing and networking resources. However, due to the underlying security and privacy threats, it is challenging to fairly motivate PVs for resource sharing in an efficient and secure way. In this paper, we propose a permissioned vehicular blockchain for secure and efficient resource sharing in VEC, namely \emph{Parkingchain}. We first design smart contract to achieve secure resource sharing and efficient service provisioning between PVs and service requesters (SRs). A multi-weight subjective logic based delegated Byzantine Fault Tolerance (DBFT) consensus mechanism is presented to improve the consensus process in Parkingchain. Further, we design a contract theory-based incentive mechanism to model the interactions between SR and PVs under asymmetric information scenario. Finally, numerical results demonstrate that the proposed incentive mechanism is effective and efficient compared with existing schemes.

\end{abstract}
\IEEEpeerreviewmaketitle
\section{Introduction}
With the rapid development of vehicular applications, a large number of connected vehicles are promoting the amount of data traffic to reach an extremely high level. For data processing, computation and communication units are installed in the vehicles \cite{han2018dynamic}. But these resources are generally underderutilized over time. According to the survey of AAA Foundation for Traffic Safety, only 50.6 minutes are spent on average driving per day by an American in 2016 \cite{han2018dynamic}. Furthermore, recent studies have shown that most of individual vehicles spend a majority of daily time on parking garage, parking lot, or driveway \cite{gu2013leverage}. In urban areas, parked vehicles (PVs) are characterized by their numerous amount and long-time fixed locations~\cite{liu2011pva}. This creates a huge opportunity to exploit the idle on-board resources (e.g. CPU and GPU) of PVs for resource sharing in vehicular networks \cite{houvehicular}, which leads to a new computing paradigm referred to as the Vehicular Edge Computing (VEC). In the existing work in the literature, PVs are utilized as static infrastructures for content downloading \cite{su2016game}, small-scale data center \cite{arif2012datacenter} and computation offloading \cite{huang2018parked}.

In the context VEC, resource sharing between service requesters (SRs) and PVs may cause security and privacy threats such as location tracking, falsification, privacy leakage, node impersonation and remote hijacking \cite{dorri2017blockchain}. Compromised PVs would heavily impede security and privacy protection for SRs via malicious behaviour. For peer-to-peer resource sharing of the PVs, SRs are generally granted access to services with necessary identity authentication. Traditionally, centralized management architectures are used to process requests and orchestrate services \cite{arif2012datacenter,gu2013leverage}. The central server becomes a bottleneck and is vulnerable to single point of failure, remote hijacking, and DDoS attacks \cite{novo2018blockchain}. Besides, if the central authority is compromised, private information of SRs and PVs may be revealed. Therefore, SRs or PVs may be not willing to upload requests or execute tasks because of the concern about data security and privacy leakage. Furthermore, some of existing studies neglect the incentive compensation for self-interested PVs which are utilized as infrastructures for computation and communication \cite{arif2012datacenter,gu2013leverage,houvehicular}. To summarize, it is necessary to design a secure and efficient incentive mechanism to motivate PVs to share their idle on-board resources in VEC.

Due to the characteristics of decentralization, anonymity and security, blockchain technology enables us to have a tamper-proof, open ledger, distributed peer-to-peer networks. The bottleneck of central server and the risks of single point of failure can be eliminated with decentralized and distributed storage and management. Blockchain has been widely studied and applied in vehicular networks, e.g. carpooling services \cite{li2018efficient}, vehicular data sharing \cite{kang2018blockchain,zhang2019data} and vehicular trust management \cite{yang2018blockchain, li2018creditcoin}. Moreover, smart contracts are self-organized scripts residing on the blockchain and allow for distributed operation of multi-step executions. In this paper, we design a smart contract that enables idle computation resource sharing for PVs in Parkingchain coupled with registration process and transaction execution to achieve the design goals.

\begin{table*}
\renewcommand{\arraystretch}{1.5}
\caption{A comparison among some blockchain systems in vehicular networks}
\begin{center}
\begin{tabu} to 1 \textwidth{p{0.15\columnwidth}<{\centering}|p{0.5\columnwidth}<{\centering}|p{0.33\columnwidth}<{\centering}|p{0.13\columnwidth}<{\centering}|p{0.3\columnwidth}<{\centering}|p{0.23\columnwidth}<{\centering}}
\hline
\textbf{Ref.}           &\textbf{Objective}       &\textbf{Network Type} &\textbf{Consensus} &\textbf{Efficiency}   &\textbf{Transaction Capacity} (tx/sec)\\
\hline
\cite{li2018creditcoin} &Vehicle communication management &Permissioned or private   &PBFT &Scalability and throughput &Thousands\\
\cite{lei2017blockchain} &Vehicle key management &Public and totally decentralized      &PoW     &Slow and high energy consumption        &Sub-ten\\
\cite{li2018efficient}  & Privacy-preserving carpooling              &Permissioned and decentralized     &PoS     &Energy efficient  &Tens\\
\cite{qiu2018blockchain}& To reach consensus in distributed software-defined vehicular networks &Private or permissioned   &PBFT  &Scalability and throughput    &Thousands\\
\cite{kang2018blockchain}&Vehicle data sharing    &Consortium and decentralized      &PoW     &Slow and high energy consumption       &Sub-ten\\
\cite{zhang2019data}    & Vehicle data sharing and storage       &Consortium and decentralized      &PBFT    &Scalability and throughput         &Thousands\\
\cite{su2018secure} &  Electric vehicle charging &Permissioned and partial centralized    &DBFT  &High scalability and high throughput   &Thousands\\
\cite{kang2019towards}   &Vehicle data sharing                       &Consortium and partial centralized     &DPoS    &Energy efficient and scalability        &Thousands\\
This paper &  Vehicle computation resource sharing &Permissioned and partial centralized    &DBFT  &High scalability and high throughput   &Thousands\\
\hline
\end{tabu}
\end{center}
\end{table*}

One issue regarding the implementation of blockchain technology is that, the most accepted proof-of-work (PoW) consensus algorithm requires an large amounts of computing power and has slow confirmation of transactions in conventional blockchain networks (e.g. Bitcoin) \cite{nakamoto2008bitcoin}. Most vehicles are resources-limited (e.g. bandwidth, computation and storage) and may not be able to meet the complex security requirements of PoW. The cost is exorbitant to build a vehicular blockchain by using computation-intensive PoW \cite{wang2018survey}. Further, traditional blockchain such as Bitcoin take up to 30 minutes for a transaction to be confirmed. However, most of vehicular applications have strict delay requirements, e.g. vehicle data sharing and computation offloading should not wait for several minutes. Therefore, PoW may not be adapted in vehicular blockchain. Alternatively, a private ledger with limited access control, fast confirmation, and low maintenance cost would be more appropriate in vehicular blockchain.

In this paper, we propose a permissioned vehicular blockchain for secure and efficient resource sharing in VEC, which we refer to as the \emph{Parkingchain}. Firstly, we design a smart contract that enables PVs' idle computation resource sharing coupled with registration process and transaction execution in Parkingchain. Secondly, a multi-weight subjective logic-based delegated Byzantine Fault Tolerates (DBFT) consensus mechanism is presented to improve the consensus process in Parkingchain. The PVs with high reputation value are selected as consensus nodes to audit and store the transaction records in Parkingchain. Finally, we propose a contract theory-based incentive mechanism to model the transactons between the SR and PVs under asymmetric information scenario. The optimal contracts are analyzed and solved to motivate PVs with different parking behaviour and energy cost while maximizing the utility of the SR.

The main contributions of this paper can be summarized as follows.
\begin{itemize}

\item We design a permissioned vehicular blockchain, referred to as \emph{Parkingchain}, where the PVs can share their idle computation resources with efficiency and security guarantee through smart contract operations. A multi-weight subjective logic-based DBFT consensus mechanism is presented to enhance the consensus process within the permissioned vehicular blockchain context.

\item We propose a contract theory-based incentive scheme to model the decision process between the SR and PVs under the asymmetric information scenario. Within the proposed framework, optimal contracts are designed to reward the PVs with different parking behaviour and energy cost for computation resource sharing while maximizing the SR's utility.

\item We present a Lagrange multiplier method-based iterative algorithm to address the optimal contract decision problem. Numerical results demonstrate that the proposed contract theory-based incentive scheme is effective and superior to other traditional schemes in terms of the utilities of SR and PVs.

\end{itemize}

The rest of this paper is organized as follows. Related work is presented in Section II. Section III introduces the system model with smart contract operations in Parkingchain. The improved DBFT consensus mechanism is presented in Section IV. In Section V, we describe contract formulation to provide incentive for the PVs and discuss the solution. Performance evaluation results are presented in Section VI before the paper is concluded in Section VII.
\begin{figure*}[t]\centering
  \includegraphics[width=1\textwidth,height=8.5cm]{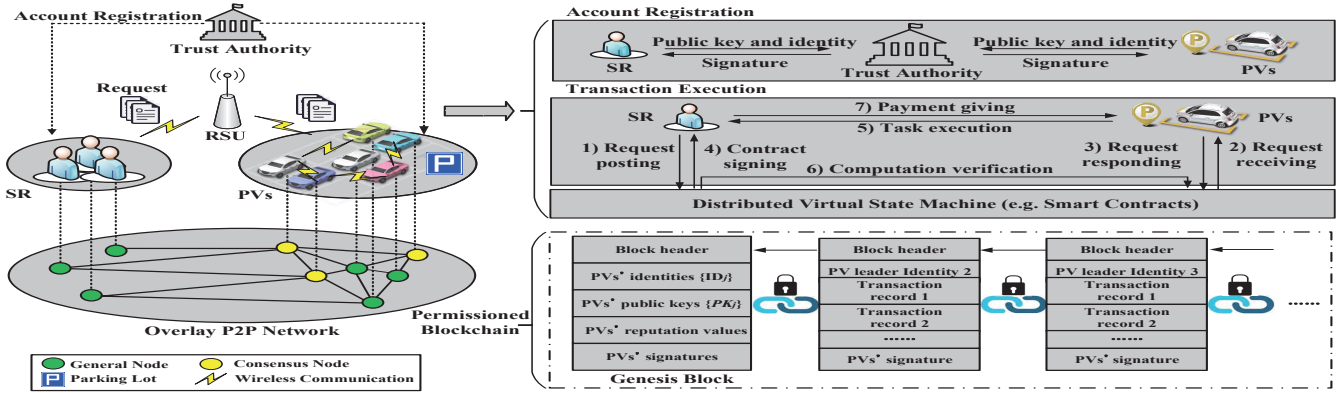}
  \caption{System model of Parkingchain.}
  \label{a}
\end{figure*}
\section{Related Work}
Recent studies have proposed different consensus algorithms to build vehicular blockchain. Table~I compares some blockchain systems in vehicular networks. We focus on those existing vehicular blockchain from five aspects: network type, consensus algorithm, efficiency, and transaction capacity. The access control decides whether the node is permitted to join blockchain, and the blockchain can be classified into permissioned type (e.g. private and consortium blockchain) and permissionless type (e.g. public blockchain). Any node in permissionless blockchain (e.g. Bitcoin \cite{nakamoto2008bitcoin}) is allowed to achieve the consensus process. On the contrast, the permissioned blockchain (e.g. Hyperledger fabric \cite{androulaki2018hyperledger}) needs a group of authenticated nodes to carry out the consensus process. The transaction capacity statistics come from \cite{xiao2019survey}, which refers to the protocol white papers and the official documentations. The transaction capacity statistics only specify the scales of magnitude, i.e. sub-ten, tens, hundreds, or thousands. Note that, network size of the BFT to process thousands of transactions per second should be small, typically around several hundred.

The most accepted PoW consensus algorithm is based on the fact that if a node performs a lot of work for the network, it is less likely that it is going to attack it. However, it requires miners to perform computationally expensive tasks and consume a large amount energy. Proof-of-stake (PoS) is similar to PoW, but the target value depends on the owned stakes of nodes. PoS is a consensus mechanism that requires less computational power than PoW, so it consumes less energy. Different from PoW and PoS applied in the public blockchain, Practical Byzantine Fault Tolerance (PBFT) \cite{castro1999practical} is carried on by a group of authenticated nodes to validate the transactions. Some variants, such as delegated PoS (DPoS) and DBFT select some nodes to generate and validate blocks to achieve high scalability, high throughput and low latency \cite{fernandez2018review}. In the proposed Parkingchain, the consensus algorithm converges to an agreement fast and is superior to a computational-based algorithm (i.e. PoW algorithm). We design our consensus phases based on DBFT algorithm to satisfy the requirements of efficiency in a vehicular networking scenario.

\section{System Model of Parkingchain}
\subsection{Network Entities in Parkingchain}
Traditional blockchain, such as Bitcoin, is a public and permissionless network. Anyone is permitted to join the blockchain, finish transactions and carry out the consensus process. In contrast, permissioned blockchain is a closed and monitored system where access privilege is well defined and constrained based on some given roles. As shown in Fig. 1, Parkingchain is a permissioned vehicular blockchain where PVs can share their idle computational resources with SRs in a secure and efficient way. We summarize main network entities in Parkingchain as follows:

\begin{itemize}
\item \emph{Service Requester}: The potential SRs have demand for computation resources in various applications, such as big data processing, scientific research and machine learning. They acts as general nodes in the overlay P2P network of Parkingchain. General nodes can relay, transmit, and exchange ledger data in Parkingchain.

\item \emph{Parked Vehicle}: PVs spend most of time in a day in the parking garage, parking lot and driveway. With wireless devices and rechargable battery, PVs can easily communicate with other vehicles and roadside units by using existing vehicular communications, such as vehicle-to-vehicle and vehicle-to-infrastructure communication protocol. PVs have two types of roles in the overlay P2P network of Parkingchain such as consensus nodes and general nodes. In addition to directly executing tasks for SRs, some preselected PVs may act as consensus nodes to validate transaction, generate block, and build hash chain over blocks.

\item \emph{Roadside Unit}: In vehicular networks, they are deployed at the network edge and become proximal access points for SRs and PVs.

\item \emph{Trust Authority}: Generally, the authority is a government agency that administers vehicle registration. It is fully trusted by all the entities in the system and responsible for issuing public parameters and cryptographic keys to them.
\end{itemize}

\subsection{Smart Contract Design for Parkingchain}
In this subsection, we design seven steps and operation details in smart contract for secure resource sharing and service provisioning in Parkingchain. The transaction between a SR and a PV is normalized through smart contract. Smart contracts are resided on Parkingchain with a unique address and triggered to handle the transactions. Smart contracts are automatically operated distributedly by network entities in a predefined manner. As shown in Fig. 1, the steps of smart contract in Parkingchain are as follows:

\begin{itemize}
\item \emph {Step 1: Account registration.} The part is to submit a user-generated account to the trusted authority for necessary authentication and registration. Asymmetric cryptography (e.g., RSA or ECC) is utilized in the system initialization for the data integrity and unforgeability. After registration on a trust authority, e.g., a government department, $PV_j$ becomes a legitimate entity and gets its public/privacy key, certificate and account address (denoted as $PK_j$, $SK_j$, $Cert_j$ and $account_{PV_j}$, respectively). Each PV's account format is composed of identity $ID_j$, public/privacy key pair $\left( {P{K_j},S{K_j}} \right)$, certificate $Cert_j$, account address $account_{PV_j}$, and average final reputation value ${{\bar g}_j}$ which is detailedly introduced in Section IV-B.

\item \emph {Step 2: Request posting.} After registration, a SR submit a request by deploying smart contract with specific requirements, including size and required computation resource of the task, expected serving time, and contract items, e.g., resource-reward pairs $\left( {{f_j},{\pi _j}} \right)$. Here, we use a contract theory-based mechanism to model the transactions between SR and PVs. More details will be given in Sections V. The smart contract contains a set of state variables including account addresses of SR and $PV_j$ $\left( {accou{t_{SR}},accou{t_{P{V_j}}}} \right)$, transaction time, and timestamp. Moreover, the SR have to submit promised reward and a deposit from their wallet address $account_{SR}$ to the smart contract address under public supervision before posting the request. This deposit will be confiscated by Parkingchain if the SR submit a fraudulent computation offloading service.

\item{ \emph {Step 3: Request receiving.} Authenticated PVs receive the request from the SR by accessing to the smart contract. The PVs learn the request requirements and promised rewards from the smart contract. According to their own preference, the PVs will decide whether to accept the request or not. }

\item{ \emph {Step 4: Contract signing .} If $PV_j$ would like to serve the SR, $PV_j$ responds to the request and sign the contract item with its private key $SK_j$ via smart contract. Similarly, $PV_j$ are required to send a deposit to smart contract for the quality of completing the task. Crucial information of SR and $PV_j$ such as identities $\left( {{ID_{SR}},{{{ID_j}}}} \right)$, certificates $\left( {{Cert_{SR}},{{{Cert_j}}}} \right)$, public keys $\left( {P{K_{SR}},{{P{K_j}}}} \right)$, account address $\left( {accou{t_{SR}},accou{t_{P{V_j}}}} \right)$, and contract item (e.g. resource-reward pairs $\left( {{f_j},{\pi _j}} \right)$) are also recorded in the smart contract. To get the reward from the SR, $PV_j$ have to complete the task and feedback computation result which can be verified by consensus nodes.}

\item{ \emph {Step 5: Task execution.} After the SR and $PV_j$ make an agreement on the contract items and sign with their private keys via smart contract, respectively. The task from will be transmitted to $PV_j$ for task execution. After receiving the task, the $PV_j$ will process it and upload the computation result to the SR and the smart contract for verification. If $PV_j$ leave the parking lot and the task is interrupted and fails, $PV_j$ will not receive any reward and the corresponding deposit will be sent to the SR for compensation. Such a punishment design is to ensure that malicious and misbehaving PVs can not hinder the service provision for the SR.}

\item{ \emph {Step 6: Computation verification.} The computation result will be verified by the consensus nodes. The consensus nodes are composed of preselected PVs with high reputation, which is detailedly presented in Section IV-B. The computation result should be verified whether the output result exactly match the input data and the task requirements. Once the results are unqualified, the deposit from $PV_j$ will be sent to the SR for compensation. Furthermore, the transaction information will be checked whether the SR's and $PV_j$'s accounts and signatures are legitimate, correct and satisfies the format requirements.}

\item{ \emph {Step 7: Payment giving.} After the computation verification stage, the promised rewards and the deposit from smart contract address are automatically sent to the $PV_j$'s account address $account_{PV_j}$. If the computation verification fails, the promised reward will be sent back to the SR and the smart contract will rollback. As a result, the transaction execution is actually the process of transactions that are being verified and written into the global ledger.}
\end{itemize}

\section{Improved DBFT for Parkingchain}
In Parkingchain, if a new data block is true and reliable, PVs will have a positive interaction and generate a positive rating for the consensus nodes. The positive interaction means that users believe the consensus service and the new data block is true and reliable \cite{kang2019towards}. Obviously, it is safe to reach consensus by PVs with high reputation. On the other hand, misbehaving PVs in Parkingchain may exhibit malicious behaviours or have some mistakes. For example, a compromised PV can be colluded with other malicious nodes and generate fake ratings, which have a negative influence on the consensus node selection. Further, an abnormal PV can be a faulty node due to connection error and does not respond requests from other nodes during consensus process, which may hinder the process of reaching consensus. Too many cheating ratings and faulty nodes in Parkingchain may lead to an unsafe and unstable environment. Therefore, it is necessary to design a safe and effective reputation management mechanism in Parkingchain.

Different from traditional PBFT \cite{castro2002practical}, PVs with high reputation are chosen as consensus nodes in the proposed DBFT. A multi-weight subjective logic model is used to calculate the reputation values of PVs based on historical interactions. Subjective logic is a framework for probabilistic information fusion, which operates on subjective beliefs about the world \cite{josang2001logic}. By monitoring behavior, each node's reputation value can be evaluated. Opinion is used to represent a subjective belief, and models positive, negative statements, and uncertainty. Following \cite{castro1999practical,su2018secure}, we design an improved DBFT consensus process by using a multi-weight subjective logic model which is based on local opinions from historical interactions and takes all the recommended opinions into consideration.

\subsection{Reputation Evaluation by Using Multi-Weight Subjective Logic Model}
A reputation value of a node can be evaluated by forming the opinion with subjective logic. The local opinion from node $i$ to node $j$ is denoted as $o_{i,j}$, which is a tuple and is defined as ${o_{i,j}} = \left( {{b_{i,j}},{d_{i,j}},{u_{i,j}},{a_{i,j}}} \right)$, where $b_{i,j}$, $d_{i,j}$ and $u_{i,j}$ denote the belief, disbelief, and uncertainty of node $i$ in node $j$, respectively, and $a_{i,j}$ is the base rate. Here, ${b_{i,j}},{d_{i,j}},{u_{i,j}},{a_{i,j}} \in \left[ {0,1} \right]$ and ${b_{i,j}} + {d_{i,j}} + {u_{i,j}} = 1.$ Let $g_{i,j}$ denote the reputation value from node $i$ to node $j$, which can be computed as ${g_{i,j}} = {b_{i,j}} + {u_{i,j}}{a_{i,j}}.$

Similar to \cite{huang2018software}, we consider the multi-weight subjective logic model for reputation opinion which is defined to measure the familiarity, timeliness and similarity. We describe them as follows:

\begin{itemize}
\item{ \emph {Familiarity:} The familiarity weight factor between PV $i$ and PV $j$ is denoted as $w_i^j$. Familiarity is defined as the interaction frequency between PV $i$ and PV $j$. A higher interaction frequency between nodes indicates that PV $i$ has more prior knowledge about PV $j$, and the PV $i$ evaluates the reputation value of PV $j$ more accurately and reliably. The familiarity weight factor $x_i^j$ between between PV $i$ and PV $j$ is calculated as
    \begin{equation}
       x_i^j = \frac{{{p_{i,j}}}}{{\frac{1}{{\left| M \right|}}\sum\nolimits_{m \in M} {{p_{i,m}}} }},
    \end{equation}
    where $p_{i,j}$ is the number of past interactions between PV $i$ and PV $j$. $M$ is the total number of PV $m$ interacted with a PV $i$.
}
\item{ \emph {Timeliness:}
The time when the reputation opinions $o_{i,j}$ is established is denoted as $t^j_i$. The current time slot is $t$. If the interaction between PV $i$ and PV $j$ is a recent event, PV $i$ will have a great impact on the local reputation opinion on PV $j$. If the opinion from PV $i$ to PV $j$ is a past event, the PV $i$ will have less influence on the local reputation on PV $j$. The timeliness weight factor $y_i^j$ between PV $i$ and PV $j$ is calculated with power-law distribution, which is denoted by
    \begin{equation}
    y_i^j = {\alpha _1}{\left( {t - t_i^j} \right)^{ - \alpha _2 }},
    \end{equation}
    where $\alpha _1$ and $\alpha _2$ are two parameters to represent the impact of interaction timeliness.
}

\item{ \emph {Similarity:}
The similarity weight factor $z_i^j$ between PV $i$ and PV $j$ describes how similar the PV $i$'s behavioral state is to the PV $j$'s behavioral state. According to the \cite{reis2016statistics}, PVs arriving at the same time in the parking lot will have similar parking behavior. It is shown that the earlier a vehicle is parked, the longer it will stay parked for. The similarity between PV $i$ and PV $j$ directly affects the final accuracy of the reputation value. The similarity value between PV $i$ and PV $j$ is computed as
    \begin{equation}
    z_i^j = \frac{1}{{1 + \left| {t_i^a - t_j^a} \right|}},
    \end{equation}
    where $t_i^a$ and $t_j^a$ are the arriving time of PV $i$ and PV $j$, respectively.
}
\end{itemize}

Considering the familiarity, opinions timeliness and similarity, the overall weight of the reputation segment ${o_{i,j}}$ can be computed by $w_i^j = {\gamma _1}x_i^j + {\gamma _2}y_i^j+ {\gamma _3}z_i^j$, where ${\gamma _1} + {\gamma _2} +{\gamma _3} = 1$. After being weighted, the recommended opinions are aggregated with the weights to obtain a synthetic reputation, which is computed as follow
    \begin{equation}
     \begin{array}{l}
      b_{k,j}^{syn} = \frac{1}{{\sum\nolimits_{k \ne j} {v_k^j} }}\sum\nolimits_{k \ne j} {w_k^j{b_{k,j}}}, \\
      d_{k,j}^{syn} = \frac{1}{{\sum\nolimits_{k \ne j} {v_k^j} }}\sum\nolimits_{k \ne j} {w_k^j{d_{k,j}}}, \\
      u_{k,j}^{syn} = \frac{1}{{\sum\nolimits_{k \ne j} {v_k^j} }}\sum\nolimits_{k \ne j} {w_k^j{u_{k,j}}}.
     \end{array}
    \end{equation}

After obtaining the recommended opinions ${o^{syn}_{k,j}}$, the final reputation value from node $i$ to node $j$ can be calculated by combining the local opinion $o_{i,j}$ with the recommended opinions ${o^{syn}_{k,j}}$. According to \cite{josang2001logic}, the final opinion is computed by $o_{i,j}^{fin} = {o_{i,j}} \oplus o_{i,j}^{syn},$ namely,
    \begin{equation}
     \begin{array}{l}
      b_{i,j}^{fin} = \frac{{{b_{i,j}}u_{k,j}^{syn} + b_{k,j}^{syn}{u_{i,j}}}}{{u_{k,j}^{syn} + {u_{i,j}} - u_{k,j}^{syn}{u_{i,j}}}},\\
      d_{i,j}^{fin} = \frac{{{d_{i,j}}u_{k,j}^{syn} + d_{k,j}^{syn}{u_{i,j}}}}{{u_{k,j}^{syn} + {u_{i,j}} - u_{k,j}^{syn}{u_{i,j}}}},\\
      u_{i,j}^{fin} = \frac{{u_{i,j}^{syn}{u_{i,j}}}}{{u_{k,j}^{syn} + {u_{i,j}} - u_{k,j}^{syn}{u_{i,j}}}}.
     \end{array}
    \end{equation}
The final reputation value from PV $i$ to PV $j$ is computed as $g_{i,j}^{fin} = b_{i,j}^{fin} + u_{i,j}^{fin}a_{i,j}^{fin}.$

\begin{figure}[t]\centering
  \includegraphics[width=0.5\textwidth,height=6cm]{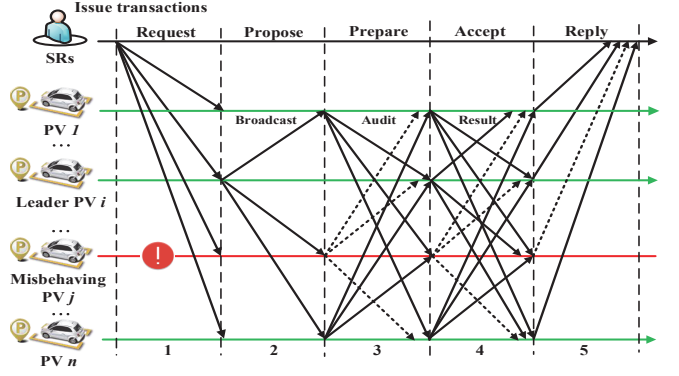}
  \caption{The consensus process in Parkingchain.}
  \label{a}
\end{figure}

\subsection{Improved DBFT Consensus Process}
\begin{itemize}

\item{ \emph {Step 1: Leader selection stage:} PVs with high reputation are chosen as consensus nodes whose reputation values are calculated with multi-weighted subjective logic model. The average final reputation values of PVs are calculated by ${{\bar g}_j} = {{\sum\nolimits_{i = 1}^M {g_{i,j}^{fin}} } \mathord{\left/{\vphantom {{\sum\nolimits_{i = 1}^M {g_{i,j}^{fin}} } M}} \right.\kern-\nulldelimiterspace} M}$, the top $n$ PVs with highest average final reputation value are selected as consensus nodes. We assume that $n> 3l+1$, where $l$ is the maximum number of malicious nodes in Parkingchain. The $n$ consensus nodes take turn to act as leader in the consensus process during $n$ time slots. Each consensus node in Parkingchain is authorized to perform block generation, broadcasting, verification and management during its turn.
    }

\item{ \emph {Step 2: Request stage:} Requesting SRs send their requests to all consensus nodes and activate the leader for service operation. All transactions are aggregated into a new data block by all consensus nodes. In addition to the leader, the other consensus nodes are called replica nodes. Firstly, the leader verifies message authentication codes (MACs) and signatures of all transactions.}

\item{ \emph {Step 3: Propose stage:} As illustrated in Figure 2, only the leader is responsible for broadcasting the new data block to replica nodes at the proposal (pre-prepare) stage. After receiving the new data block from leader, each replica nodes verifies MACs and signatures of all transactions in the new data block.

}

\item{ \emph {Step 4: Prepare stage:} The verified results are added to replica nodes' digital signatures and are broadcast to all others. After receiving $2l$ messages from different nodes, the prepare stage will be finished and it will enter the following steps.}

\item{ \emph {Step 5: Accept stage:} After receiving the audit results from other replica nodes, all consensus nodes compare them with their own audit results. Then all consensus nodes broadcast an acknowledgment message to all others. If more than $n-l$ consensus nodes agree on the data block, it replies to the SR and writes the result to the block. }

\item{ \emph {Step 6: Reply stage:} If the consensus nodes and the leader receive a certain number of proposal approvals from the others, they will reply to the SR. If the number of abnormal nodes is less than $\left( {n - l} \right)/3$, the consensus result can be guaranteed correctly. }
\end{itemize}

These classical Byzantine Faulty Tolerant (BFT)-based consensus schemes ensure the properties of deterministic approval and liveness in Byzantine environment, and are characterized by low processing delay \cite{vukolic2015quest}. The characteristics of leader-replica hierarchy has message complexity of ${O}\left( {{n^2}} \right)$ \cite{cachin2009yet}. Therefore, the BFT-based consensus approaches should be applied in a small-scale permissioned network with centralized admission control \cite{wang2018survey}.

\section{Contract Theory Based Optimization For Parkingchain}
In this section, we aim to design an incentive scheme to motivate PVs to share their idle computation resource in Parkingchain. Due to the energy cost incurred by task execution, SR should pay for the computation offloading services provided by PVs. Nevertheless, there is another issue to be solved in designing the incentive approach. Obviously, PVs may leave parking place at any time. The link and the computation offloading service will be interrupted when the PVs leave parking place occasionally. It is worth noting that the parking behaviour of PVs is the private information, which is known by the PVs themselves but unavailable for the SR. The SR is not aware of the parking behaviour of PVs, which leads to the information asymmetry between the SR and PVs. The asymmetric information causes additional cost for SR to employ PVs to help offload task. Therefore, it is necessary to design an incentive scheme which can overcome the asymmetric information scenario effectively.

Considering that contract theory is a powerful tool from economics to handle incentive problem under asymmetric information scenario, we leverage contract theory to design an incentive scheme to motivate the PVs in Parkingchain. The SR is modeled as the employer who offers a contract to each PV. Each contract is composed of a serious of contract items, which are combinations of resource-reward pairs.
\subsection{PVs Type Modeling}
A vehicle's parking behavior can be described a dual Gamma distribution~\cite{reis2016statistics}. Let $t_p$ denote the time that is spent parked by a PV. The time when the vehicle arrives at parking lot is denoted as $t_a$. The first-order density of $t_p$ is given by
\begin{equation}
\begin{aligned}
f\left( {t_p,{t_a}} \right) &= \frac{{H_{{t_a}}^s}}{{\Gamma \left( {\kappa _{{t_a}}^s} \right){{\left( {\varepsilon _{{t_a}}^s} \right)}^{\kappa _{{t_a}}^s}}}}{t_p^{\kappa _{{t_a}}^s - 1}}{e^{ - \frac{t_p}{{\varepsilon _{{t_a}}^s}}}} \\ &+ \frac{{H_{{t_a}}^l}}{{\Gamma \left( {\kappa _{{t_a}}^l} \right){{\left( {\varepsilon _{{t_a}}^l} \right)}^{\kappa _{{t_a}}^l}}}}{t_p^{\kappa _{{t_a}}^l - 1}}{e^{ - \frac{t_p}{{\varepsilon _{{t_a}}^l}}}},
\end{aligned}
\end{equation}
where $t_p > 0$, and $t_a = \left\{ {0,1,2,...,23} \right\}$. The coefficients ${\kappa _{{t_a}}^s},{\varepsilon _{{t_a}}^s}$ are the shape and scale parameters that model short-term parking behavior of PVs, while the coefficients ${\kappa _{{t_a}}^l},{\varepsilon _{{t_a}}^l}$ are the parameters that models long-term parking behavior of PVs. The coefficients ${H_{{t_a}}^s}$ and ${H_{{t_a}}^l}$ must always integrate to one (e.g. ${H_{{t_a}}^s}+{H_{{t_a}}^l}=1$). $\Gamma \left(  \cdot  \right)$ is the Gamma function. The values of these coefficients can be referred to \cite{reis2016statistics}.

The proposed blockchain network runs on a slotted-time basis with the time axis divided into the equal non-overlaping time slots (Hour) of a day. Let $t$ denote an integer-valued slot index, $t_i^{p}, i \in \left\{ {1,...,M} \right\}$ denote the time a PV has spent parked, and $t_i^a\left( t \right)=t-t_i^p$ denote the arrival time of a PV. According to \cite{reis2016statistics,sun2018end}, the probability that a PV has at least $\tau$ hours left for parking when it has been parked for $t_i^p$, $i \in \left\{ {1,...,M} \right\}$ at time $t$ is given by
\begin{equation}
\begin{array}{l}
P_i^s\left( t \right) = P\left[ {t > t_i^p + \tau \left| {t > t_i^p} \right.} \right] = \\
\frac{{H_{t_i^a}^s\gamma \left( {\kappa _{t_i^a}^s,\frac{{t_i^p + \tau }}{{\varepsilon _{t_i^a}^s}}} \right)\Gamma \left( {\kappa _{t_i^a}^s} \right) + H_{t_i^a}^l\gamma \left( {\kappa _{t_i^a}^l,\frac{{t_i^p + \tau }}{{\varepsilon _{t_i^a}^l}}} \right)\Gamma \left( {\kappa _{t_i^a}^l} \right) - \Gamma \left( {\kappa _{t_i^a}^l} \right)\Gamma \left( {\kappa _{t_i^a}^s} \right)}}{{H_{t_i^a}^s\gamma \left( {\kappa _{t_i^a}^s,\frac{{t_i^p}}{{\varepsilon _{t_i^a}^s}}} \right)\Gamma \left( {\kappa _{t_i^a}^s} \right) + H_{t_i^a}^l\gamma \left( {\kappa _{t_i^a}^l,\frac{{t_i^p}}{{\varepsilon _{t_i^a}^l}}} \right)\Gamma \left( {\kappa _{t_i^a}^l} \right) - \Gamma \left( {\kappa _{t_i^a}^l} \right)\Gamma \left( {\kappa _{t_i^a}^s} \right)}}.
\end{array}
\end{equation}

Obviously, the PVs may leave parking place at any time. The link and the computation offloading services of the PVs will be interrupted when the PVs leave parking place. The probability that a PV leaves after it has been parked for $t_i^p$ time, can be expressed by
\begin{equation}
\begin{aligned}
P_i^o\left( t \right) &= P\left[ {t \le t_i^p + \tau \left| {t > t_i^p} \right.} \right] \\&= 1 - {P_i}\left[ {t > {t_i^p} + \tau \left| {t > {t_i^p}} \right.} \right] = 1 - P_i^s\left( t \right).
\end{aligned}
\end{equation}

The SR is not aware of the parking behaviour of PVs, which leads to asymmetric information between the SR and the PVs. The SR can classify the PVs into some discrete types and use the statistical distributions of the PVs' parking behaviour from historical data to improve its own benefit.

\noindent\textbf{Definition 1:} We define the $j$-th type of PVs ${\theta _j}\left( t \right)$ at time $t$ as the probability that the PVs has at least $\tau$ hours left for parking when it has been parked for $t_j^p$, $j \in \left\{ {1,...,N} \right\}$
\begin{equation}
{\theta _j}\left( t \right) = P_j^s\left( t \right) = 1 - P_j^o\left( t \right).
\end{equation}
The set of PVs types at time $t$ is denoted as $\Theta \left( t \right) = \left\{ {{\theta _1}\left( t \right),{\theta _2}\left( t \right),...,{\theta _N}\left( t \right)} \right\}$. The probabilities are sorted in an ascending order and classified into $N$ types. $\Theta \left( t \right)$ denotes the types of PVs at time $t$ and  ${\theta _1}\left( t \right) < ... < {\theta _j}\left( t \right) < ... < {\theta _N}\left( t \right)$. In this definition, the higher type of PVs have a higher probability that the PVs have at least $\tau$ time left for parking when it has been parked for time $t_j^p$.

\subsection{Contract Formulation}
In order to improve the quality of experience (QoE) of SR, it is essential for PVs to guarantee the latency requirement of SR. The satisfaction function model that we use in this paper is similar to \cite{liwang2018truthful}. The physical significance of time saved if SR offloads a task to type-$j$ PVs is given by

\begin{equation}
{S_j}\left( t \right) =\rho \left( \frac{\kappa s }{{{f_{local}}}} - \frac{\kappa s}{{{f_j}\left( t \right)}} - \frac{s}{{{r_j}}}\right),
\end{equation}

\noindent where $\rho$ is profit coefficient for unit time saved, $\kappa$ is the mapping from bit to CPU cycles, $s$ is the size of a task, and $f_{local}$ denotes computation capacity that the task is processed locally. The computation resources shared by the type-$j$ PVs at time $t$ is denoted as $f_{j}\left( t \right)$, and $r_j$ denotes the transmission rate between the type-$j$ PVs and the SR. Obviously, higher values of $f_j\left( t \right)$ can provide more benefits to the SR. The SR's expected utility with type-$j$ PVs at time $t$ is given by
\begin{equation}
\begin{aligned}
U_{SR_j}\left( t \right) &=  {P_j^s\left( t \right)\left( { {S_j}\left( t \right) - {\pi _j}\left( t \right)} \right) + P_j^o\left( t \right)0} \\ &=\theta_j\left( t \right)\left( { {S_j}\left( t \right) - {\pi _j}\left( t \right)} \right),
\end{aligned}
\end{equation}
Under asymmetric information scenario, the SR only knows the probability of the PVs belonging to a certain type $\theta_j\left( t \right)$ from historical statistics. The probability that the PVs belong to type-$j$ at time $t$ can be denoted as: $\beta_j\left( t \right)$, and $\sum\nolimits_{j = 1}^N {{\beta_j}}\left( t \right)  = 1$. The expected utility of SR at time $t$ can be defined as
\begin{equation}
\begin{aligned}
{U_{SR}}\left( t \right)&= \sum\limits_{j = 1}^N {\beta_{j}}\left( t \right)\theta_j\left( t \right) \left( {S_j}\left( t \right)  -  {{\pi _{j}}\left( t \right)} \right),
\end{aligned}
\end{equation}
\noindent where $\pi_{j}\left( t \right)$ is the payment from SR for the offloading services provided by the type-$j$ PVs at time $t$.

The energy consumption of PVs is modelled by a quadratic function. Let $E_j\left( t \right)$ denote the energy consumption of the type-$j$ PVs at time $t$ and $f_{j}\left( t \right)$ represent the computation resources (i.e. the clock frequency of the CPU chip) contributed by the type-$j$ PVs at time $t$. The energy consumption of the type-$j$ PVs for offloading a task at time $t$ can be defined as
\begin{equation}
{E_j}\left( t \right) = \kappa s\varepsilon { {{f_j}\left( t \right)}^2} = \kappa s\varepsilon {f_j}{\left( t \right)^2},
\end{equation}
where $\varepsilon$ is the effective switched capacitance that depends on the chip architecture. $E_j\left( t \right)$ is non-negative and monotone increasing on $f_{j}\left( t \right)$. The enery cost of the type-$j$ PVs at time $t$ is given by
\begin{equation}
C_j\left( t \right) = eE_j\left( t \right) = e\kappa s \varepsilon {f_j}{\left( t \right)^2},
\end{equation}
where $e$ is price for energy consumption.

The expected revenue of the type-$j$ PVs at time $t$ is
\begin{equation}
R_j\left( t \right) = P_j^s\left( t \right) v\left( {{\pi _j}}\left( t \right) \right) + P_j^o\left( t \right) 0  = {\theta _j}\left( t \right)v\left( {{\pi _j}}\left( t \right) \right),
\end{equation}
where $v\left( {{\pi _j}}\left( t \right) \right)$ is the evaluation function regarding the reward ${{\pi _j}}\left( t \right)$ paid by the SR, which is a strictly increasing concave function of ${{\pi _j}}\left( t \right)$, and $v\left( 0 \right)=0$, $v\left( {{\pi _j}}\left( t \right) \right)^\prime  > 0$, and $v\left( {{\pi _j}}\left( t \right) \right)^{\prime \prime } < 0$ for all ${{\pi _j}}\left( t \right)$, $j \in \left\{ {1,...,N} \right\}$. For $j \in \left\{ {1,...,N} \right\}$, the expected utility of the type-$j$ PVs under contract $\left( {f_j}\left( t \right),{{\pi _j}}\left( t \right) \right)$ at time $t$ takes the form of
\begin{equation}
U_{PV_j} \left( t \right)= R_j\left( t \right) - C_j\left( t \right) = {\theta _j}\left( t \right)v\left( {{\pi _j}}\left( t \right) \right) - e\kappa s \varepsilon  {f_j}{\left( t \right)^2}.
\end{equation}
The expected utility of PVs is the expected revenue minus the energy cost of task execution.

To simplify the notations, we skip the index $t$ in all the variables below. The problem of maximizing the expected utility of the SR (contract design optimization) is formulated as
\begin{equation}
\begin{aligned}
&\mathop {\max }\limits_{\left( {{f_j},{\pi _j}} \right)} \;\sum\limits_{j = 1}^N {\beta_j}\theta_j \left[  {\rho \left( {\frac{\kappa s}{{{f_{local}}}} - \frac{\kappa s}{{{f_{j}}}}} - \frac{s}{{{r_j}}}\right) -  {{\pi _{j}}}} \right] \\
&\;\;s.t. \; \,(17a) \;  {\theta _j}v\left( {{\pi _j}} \right) - e\kappa s \varepsilon {f_j}^2 \ge {\theta _j}v\left( {{\pi _k}} \right) - e\kappa s \varepsilon {f_k}^2, \\ & \quad\quad\quad\quad\;\; \forall j \in \left\{ {1,...,N} \right\}, \\
&\;\;\;\; \; \quad  (17b) \;  {\theta _j}v\left( {{\pi _j}} \right) - e\kappa s \varepsilon {f_j}^2 \ge 0,\forall j \in \left\{ {1,...,N} \right\},\\
&\;\;\;\; \; \quad  (17c)\;  0 \le {f_1} \le {f_2} \le ... \le {f_{\max }}.   \\
\end{aligned}
\label{contract-opt}
\end{equation}
The incentive compatibility (IC) constraint in (\ref{contract-opt}a) guarantees that the type-$j$ PV can reach the maximum value by selecting the contract item $(\pi_j,f_j)$. Besides, the individual rationality (IR) constraint in (\ref{contract-opt}b) guarantees that the utility of each type of PVs is non-negative. To encourage the PVs to share their computation resources, the reward given by the SR for each type of PVs must compensate their energy cost. Obviously, the optimization problem in (\ref{contract-opt}) is not a convex optimization problem. We will simplify the  constraints in the optimization problem as following.

\subsection{Problem Transformation}
It is difficult to solve the optimization problem (17) with the complicated constraints including $N\left(N-1\right)$ IC constraints and $N$ IR constraints, which are not-convex and coupled among different types of PVs \cite{zhang2015contract}. The constraints of (\ref{contract-opt}) will be reduced through the following lemmas.

\noindent \textbf{Lemma 1:} \emph{(Reduce the IR constraints) If the utility of the SR is maximized under asymmetric information scenario, given that the IC constraints are satisfied, the IR constraints can be replaced by}
\begin{equation}
{\theta _1}v\left( {{\pi _1}} \right) - e\kappa s \varepsilon {f_1}^2 = 0.
\end{equation}

\textbf{\emph{Proof:}} See \textbf{Appendix A}.              \hfill $\blacksquare$

\noindent \textbf{Lemma 2:} \emph{ For any feasible contract $\left( {\pi_j ,f_j } \right)$, ${\pi_j} > {\pi_k}$ \emph{if and only if} ${{\theta _j} > {\theta _k}}, \forall j,k \in \left\{ {1,...,N} \right\}$.}

\textbf{\emph{Proof:}} See \cite{bolton2005contract}.               \hfill         $\blacksquare$


\noindent \textbf{Lemma 3:} \emph{ For any feasible contract $\left( {\pi_j ,f_j } \right)$, ${\pi_j} > {\pi_k}$ \emph{if and only if} ${{f _j} > {f _k}},\forall j,k \in \left\{ {1,...,N} \right\}$.}

\textbf{\emph{Proof:}} See \cite{bolton2005contract}.\hfill         $\blacksquare$


\noindent \textbf{Lemma 4:}\emph{ (Reduce the IC constraints) Based on \textbf{Lemma 1} and \textbf{Lemma 2}, the IC constraints can be reduced as the local downward incentive compatibility (LDIC):}
\begin{equation}
\begin{aligned}
{\theta _j}v\left( {{\pi _j}} \right) & - e\kappa s\varepsilon {f_j}^2 \ge {\theta _j}\pi_{j-1} - e\kappa s \varepsilon {f_{j-1}}^2,\\
&{\forall j \in \left\{ {2,...,N} \right\}},
\end{aligned}
\end{equation}
\noindent and the local upward incentive compatibility (LUIC):
\begin{equation}
\begin{aligned}
{\theta _j}v\left( {{\pi _j}} \right) & - e\kappa s \varepsilon {f_j}^2 \ge {\theta _j}\pi_{j+1}- e\kappa s\varepsilon {f_{j+1}}^2,\\
&{\forall j \in \left\{ {1,...,N-1} \right\}},
\end{aligned}
\end{equation}

\textbf{\emph{Proof:}} See \textbf{Appendix D}.        \hfill $\blacksquare$

\noindent \textbf{Lemma 5:} \emph{ If the utility of SR is maximized, the IC constraints can be reduced as}
\begin{equation}
{\theta _j}v\left( {{\pi _j}} \right) - e\kappa s \varepsilon {f_j}^2 = {\theta _j}v\left( {{\pi _{j-1}}} \right) - e\kappa s \varepsilon {f_{j-1}}^2.
\end{equation}

\textbf{\emph{Proof:}} See \textbf{Appendix E}.        \hfill $\blacksquare$

Based on \textbf{Lemma 1} to \textbf{Lemma 5}, the complicated IR and IC constraints can be reduced. Therefore, the optimization problem in (\ref{contract-opt}) can be further simplified as follows
\begin{equation}
\begin{aligned}
&\mathop {\max }\limits_{\left( {{f_j},{\pi _j}} \right)} \;\sum\limits_{j = 1}^N {{\beta _j}} {\theta _j}\left[ {\rho \left( {\frac{{\kappa s}}{{{f_{local}}}} - \frac{{\kappa s}}{{{}{f_j}}} - \frac{s}{{{r_j}}}} \right) - {\pi _j}} \right] \\
&\;\;s.t. \; \,(22a) \;  {\theta _j}v\left( {{\pi _j}} \right) - e\kappa s\varepsilon  {f_j}^2 = {\theta _j}v\left( {{\pi _{j-1}}} \right) - e\kappa s \varepsilon  {f_{j-1}}^2 , \\ & \quad\quad\quad\quad\;\; \forall j \in \left\{ {2,...,N} \right\}, \\
&\;\;\;\; \; \quad  (22b) \;  {\theta _1}v\left( {{\pi _1}} \right) - e\kappa s\varepsilon  {f_1}^2  = 0,\\
&\;\;\;\; \; \quad  (22c)\; 0 \le {f_1} \le {f_2} \le ... \le {f_{\max }}.
\end{aligned}
\label{opt-simplified}
\end{equation}

The problem in (\ref{opt-simplified}) can be easily proved to be a convex programming problem by checking the convexity of the objective function and the constraints. Thus, problem (\ref{opt-simplified}) can be solved for the global optimal solution by using the Lagrange multiplier method.

\subsection{Solution to Optimal Contracts}
\subsubsection{Optimal Contracts for Complete Information Scenario}
Under complete information scenario, the SR is ideally aware of all PVs' type. Given that, the SR only has to provide acceptable contract to the PVs. The SR tries to extract all the revenue from the PVs to maximize its own utility, which leaves PVs zero utilities. We set $U_{PV_j}=0, \forall j \in \{1,...,N\}$ and we have,
 \begin{equation}
\begin{array}{*{20}{c}}
{{f_j} = }
\end{array} \sqrt {\frac{{{\theta _j}v\left( {{\pi _j}} \right)}}{{e\kappa s\varepsilon }}} ,\forall j \in \left\{ {1,...,N} \right\}.
\end{equation}
Then we can replace the $f_j$ in $U_{SR_j}$ with (23) and the IC constraints can be neglected under complete information scenario. We can obtain the optimal contract $\left\{ {\bar f_j,\bar \pi _j} \right\}, \forall j\in \{1,...,N\}$ for each type of PVs by solving ${{d{U_{S{R_j}}}} \mathord{\left/{\vphantom {{d{U_{S{R_j}}}} {d{\pi _j}}}} \right.\kern-\nulldelimiterspace} {d{\pi _j}}} = 0$. We consider the complete information scenario as a benchmark case, and take the optimal SR’s expected utility under complete information scenario as the up-bound for the performance.

\textbf{Lemma 6:} \emph{ Under complete information scenario, the optimal ${{\bar \pi }_j}, \forall j\in \{1,...,N\}$ can be obtained by solving ${{d{U_{S{R_j}}}} \mathord{\left/{\vphantom {{d{U_{S{Q_j}}}} {d{\pi _j}}}} \right.\kern-\nulldelimiterspace} {d{\pi _j}}} = 0, \forall j\in \{1,...,N\}.$} \\
\textbf{\emph{Proof:}} See \textbf{Appendix F}.     \hfill $\blacksquare$

\subsubsection{Local Optimal Contracts for Asymmetric Information Scenario}
Based on constraints (\ref{opt-simplified}a) and (\ref{opt-simplified}b), we have equations as follow
\begin{equation}
{f_j} = \left\{ {\begin{array}{*{20}{c}}
{\sqrt {\frac{{{\theta _1}v\left( {{\pi _1}} \right)}}{{e\kappa s\varepsilon }}} ,}&{j = 1,}\\
{\sqrt {\frac{{{\theta _1}v\left( {{\pi _1}} \right) + \sum\nolimits_{k = 2}^j {{\Delta _k}} }}{{e\kappa s\varepsilon }}} ,}&{2 \le j \le N},
\end{array}} \right.
\end{equation}
where ${\Delta _k} = {\theta _k}\left[ {v\left( {{\pi _k}} \right) - v\left( {{\pi _{k - 1}}} \right)} \right]$. To solve objective function (\ref{opt-simplified}), we replace $f_j$ in (\ref{opt-simplified}) with (24). We can obtain the optimal contract $\left\{ {\dot f_j,\dot \pi _j} \right\}, \forall j\in \{1,...,N\}$ for each type of PVs by solving ${{d{U_{S{R_j}}}} \mathord{\left/{\vphantom {{d{U_{S{R_j}}}} {d{\pi _j}}}} \right.\kern-\nulldelimiterspace} {d{\pi _j}}} = 0$. However, $\dot \pi _j, \forall j\in \{1,...,N\}$ are not global optimal, because the solution $\dot \pi _j$ depends on the other solution $\dot \pi _{j-1}$.

\noindent \textbf{Lemma 7:} \emph{ Under asymmetric information scenario, the local optimal ${{\dot \pi }_j}, \forall j\in \{1,...,N\}$ can be obtained by solving ${{d{U_{S{R_j}}}} \mathord{\left/{\vphantom {{d{U_{S{R_j}}}} {d{\pi _j}}}} \right.\kern-\nulldelimiterspace} {d{\pi _j}}} = 0, \forall j\in \{1,...,N\}.$} \\
\textbf{\emph{Proof:}} See \textbf{Appendix G}.     \hfill $\blacksquare$

\subsubsection{Lagrange Multiplier Method-Based Iterative Algorithm}
Under asymmetric information scenario, we propose a Lagrange multiplier method based iterative algorithm to solve the optimization problem in (\ref{opt-simplified}) without constraint (\ref{opt-simplified}c). After that, we need to check whether the solution of the relaxed problem satisfies this constraint. Based on the optimization problem (\ref{opt-simplified}), the Lagrangian can be written as
\begin{equation}
\begin{aligned}
 {\cal L} &= \sum\limits_{j = 1}^N {\left\{ {{\beta _j}{\theta _j}\left[ {\rho \left( {\frac{{\kappa s}}{{{f_{local}}}} - \frac{{\kappa s}}{{{}{f_j}}} - \frac{s}{{{r_j}}}} \right) - {\pi _j}} \right]} \right\}}  \\
& +\sum\limits_{j = 2}^N {\omega _j}\left\{ {{\theta _j}\left[ {v\left( {{\pi _j}} \right) -v\left( {{\pi _{j-1}}} \right)} \right] - e\varepsilon\kappa s  \left( {{f_j}^2 - f_{j - 1}^2} \right)} \right\} \\
& + \eta \left\{ {{\theta _1}\left(\pi _{1}\right) - e\kappa s\varepsilon {f_1}^2} \right\}.
\end{aligned}
\end{equation}
\begin{algorithm}[t]
\caption{Contract Optimization Based on Lagrange Multiplier Method}
\textbf {Input:} $\left\{ {{\theta _{j}}} \right\}$, $\left\{ {{\beta _{j}}} \right\}$, $\{r_j\}$, $\rho$, $f_{local}$, $\kappa$, $e$, $\varepsilon$ \\
\textbf {Output:} The optimal contract $\left\{ {f_{j}^*,\pi _{j}^*} \right\},\forall j \in \left\{ {1,...,N} \right\}$.   \\
Compute $\bar \pi_N$ and $\dot \pi_N$ by solving ${{d{U_{S{R_j}}}} \mathord{\left/{\vphantom {{d{U_{S{Q_j}}}} {d{\pi _j}}}} \right.\kern-\nulldelimiterspace} {d{\pi _j}}} = 0$ under complete information scenario and asymmetric information scenario, respectively; \\
Then we obtain $\bar f_N$ and obtain $\dot f_N$ with (23) and (24), respectively. Set $f_N^*=\bar f_N$, $\delta  = \frac{{{{\bar f }_N} - {{\dot f }_N}}}{S}$, $j=N$; \\
\While{$j \ne 1$}
{
  Calculate $\omega _N^* = \frac{{{\beta _N}{\theta _N}\rho }}{{2e\varepsilon \mu _p^3 f_N^* }}$ with (28) and $\pi _N^* = \frac{{\omega _N^*}}{{{\beta _N}}} - 1$ with (29) \\
  \For {$j \leftarrow N - 1$ to $1$}
{
  Calculate optimal contracts $\left\{ f_{{j}}^*,\pi _{{j}}^* \right\}$ with (26), (27) and (30)); \\
    Calculate $U_{PV_j}^*$ with (11); \\
   \If {$U_{PV_j}^*<0$}
  {
  \textbf{Break for}
  }
}
  ${f_N^*} \leftarrow {f_N^*} - \delta $
}
\While {there exists an infeasible sub-sequence $\{f_j^*\}$}
{
 Find an infeasible sub-sequence $\{f_m^*,f_{m+1}^*,...,f_n^*\}$ \\
 Set $f_j^* = \arg {\max _{\left\{ f \right\}}}\sum\nolimits_{k = m}^n {{U_{S{R_k}}}} ,\forall j \in \left\{ {m,m + 1,...,n} \right\}$; \\
 \For {$j \leftarrow 1$ to $N$}
 {
 Assign the optimal price $\pi_j^*$ based on (24);
 }
}
\end{algorithm}
\noindent The Lagrange multiplier associated with the simplified IC constraints for type $\theta_j$ is $\omega_j$, while $\eta$ is the multiplier associated with the simplified IR constraint for type-$1$. For $j \in \left\{ {1,...,N-1} \right\}$, the first-order conditions are given by
\begin{equation}
\frac{{\partial L}}{{\partial {f_j}}} = \frac{{{\beta _j}{\theta _j}\rho \kappa s}}{{{}{f_j}^2}} - 2e\varepsilon \kappa s {f_j}\left( {{\omega _j} - {\omega _{j + 1}}} \right) = 0,
\end{equation}
\begin{equation}
\frac{{\partial {\cal L}}}{{\partial {\pi _j}}} =  - {\beta _j}{\theta _j} + v'\left( {{\pi _j}} \right)\left( {{\omega _j}{\theta _j} - {\omega _{j + 1}}{\theta _{j + 1}}} \right) = 0,
\end{equation}
The partial derivatives regarding $\pi_j$ and $f_j$ when $j=N$ are
\begin{equation}
\frac{{\partial L}}{{\partial {f_N}}} = \frac{{{\beta _N}{\theta _N}\rho \kappa s}}{{{}{f_N}^2}} - 2e\varepsilon\kappa s  {\omega _N}{f_N} = 0,
\label{eq-28}
\end{equation}
\begin{equation}
\frac{{\partial {\cal L}}}{{\partial {\pi _N}}} =  - {\beta _N}{\theta _N} + v'\left( {{\pi _N}} \right){\omega _N}{\theta _N} = 0.
\label{eq-29}
\end{equation}
With constraints (22a), we have
\begin{equation}
e\varepsilon\kappa s f_j^2 = {\theta _j}\left( {v\left( {{\pi _j}} \right) - v\left( {{\pi _{j - 1}}} \right)} \right) + e\varepsilon\kappa s f_{j - 1}^2.
\end{equation}

To solve the optimal contracts $\left\{ {f_j^*,\pi _j^*} \right\}$, we start the calculation with the highest type of PVs $\theta_N$. If $f_N^*$ is given, we obtain $\omega_N^*$ and $\pi_N^*$ according to (\ref{eq-28}) and (\ref{eq-29}). Then, the optimal contracts $\left\{ {f_j^*,\pi _j^*} \right\}$ for all $j\in N$ can be calculated according to (26), (27) and (30). Hence, we develop Algorithm 1 to obtain the optimal solutions by searching the optimal $f_N^*$ iteratively. However, the obtained set $\{f_j^*\}$ may not satisfy the constraint (22c). The sub-sequences of the set, which are not in the increasing order, are called infeasible sub-sequences \cite{gao2011spectrum}. Since $\{U_{SR_j}\}$ are concave function, the infeasible sub-sequences can be replaced by feasible sub-sequences iteratively \cite{gao2011spectrum}. The details are shown in \textbf{Algorithm 1}.



\begin{table}[t]
\renewcommand{\arraystretch}{1}
\caption{Simulation parameters}\label{table} \centering \tabcolsep=5pt
\begin{tabular}{p{4.3cm}|p{3.5cm}}
\hline
\textbf{Parameter} & \textbf{Setting} \\
\hline

Number of misbehaving PVs & 10\\
\hline
Interaction frequency between PVs, $p$ & $U \sim \left[5,10\right]$/minutes\\
\hline
Quality of link, $q$ & $U \sim \left[ {0.5,0.9} \right]$\\
\hline
Weight of reputation segment, $\gamma_1$, $\gamma_2$, and $\gamma_3$  & $0.3, 0.4, 0.3$\\
\hline
Predefined parameters of opinions timeliness, $\alpha_1$ and $\alpha_2$ & $10, 1.5$\\
\hline
Number of type, $N$ & $7$\\
\hline
Local computation capacity, ${{f_{local}}}$ & $0.5*10^9$ Hz\\
\hline
Mapping from bit to cycles, $\kappa$ & $10^4$ cycle/bit\\
\hline
Transmission rate, $r_j$ & $5-6$ Mb/s \\
\hline
Size of a task, $s$ & $500$ KB\\
\hline
Effective switched capacitance, $\varepsilon$ & $10^{-28}$\\
\hline
Profit coefficient, $\rho$ & $0.1$\\
\hline
Energy cost coefficient, $e$ & $0.1$\\
\hline
\end{tabular}\label{table1}
\end{table}

\section{Numerical Results}

\begin{figure}[t]\centering
  \includegraphics[width=0.5\textwidth,height=6.5cm]{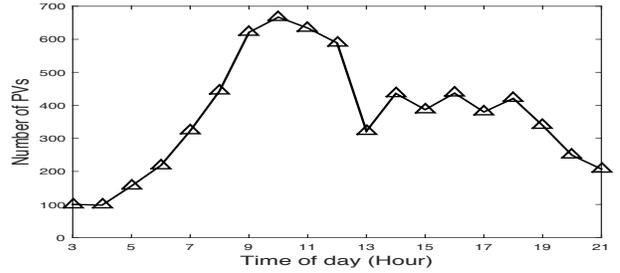}
  \caption{Number of PVs versus arrival time.}
  \label{a}
\end{figure}

\begin{figure}[t]\centering
  \includegraphics[width=0.5\textwidth,height=6.5cm]{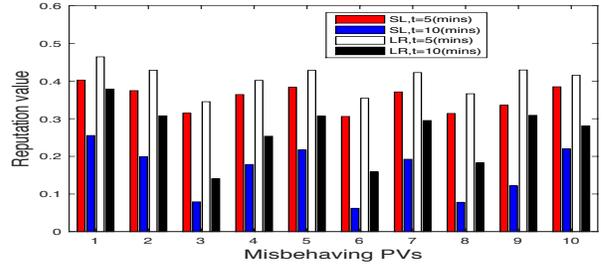}
  \caption{The reputation values of 10 misbehaving PVs.}
  \label{a}
\end{figure}
\begin{figure}[t]\centering
  \includegraphics[width=0.5\textwidth,height=6.5cm]{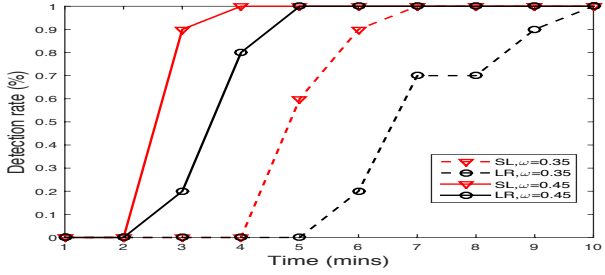}
  \caption{Detection rate of misbehaving consensus nodes over time.}
  \label{a}
\end{figure}
\begin{figure}[t]\centering
  \includegraphics[width=0.5\textwidth,height=6.5cm]{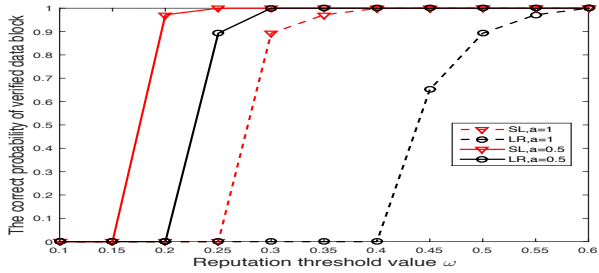}
  \caption{Probability of corrected blocks versus different reputation threshold value.}
  \label{a}
\end{figure}
\begin{figure}[t]\centering
  \includegraphics[width=0.5\textwidth,height=6.5cm]{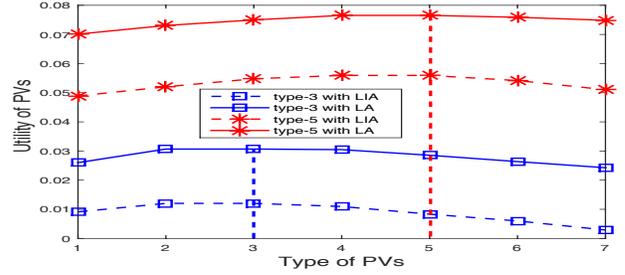}
  \caption{Utility of PVs under two schemes with different type of PVs.}
  \label{a}
\end{figure}

Firstly, we evaluate the performance of the proposed Subjective logic based scheme (SL) in DBFT consensus process. We introduce another traditional linear reputation scheme (LR) for comparison \cite{huang2018software}. Secondly, we give an analysis about the feasibility of the proposed contract. Thirdly, we conduct comparisons among the utilities of the SR and the PVs by varying the time in a day. Finally, we evaluate the utilities of the SR and PVs with different schemes versus different types of PVs and In the following figures, the numerical results are obtained by solving the problem by using proposed Lagrangian-based iterative scheme under asymmetric information scenario (LIA), local optimization under complete information scenario (LC) and asymmetric information (LA). For comparisons, we introduce another two incentive schemes. The first one is the Stackelberg game scheme under asymmetric information scenario (SA) \cite{liu2017design}. The second one is the linear pricing scheme \cite{zhang2016offloading}. Without loss of generality, we consider $v\left( {{\pi _j}} \right) = \ln \left( {1 + {\pi _j}} \right),\forall j \in \left\{ {1,2,...,N} \right\}$. The system performance is simulated using MATLAB with system parameters in TABLE~II.

We adopt the real traces from ACT Government Open Data Portal dataACT \cite{dataset}. The data set is from SmartParking app which is a trial designed to help ease traffic congestion and lower travel times by using real-time bay sensor data. This trace was presented from 180295 parking records in the Manuka shopping precinct in a month. The probability distribution for the different arrival time of PVs is shown in Fig. 3. The range of the arrival time is from $3$ A.M to $9$ P.M. Based on the number of PVs with different arrival time, the probability that the PVs belong to each type at time $t$ can be calculated.

\subsection{Performance of the Proposed Subjective Logic-Based Reputation Mechanism}
In Fig. 4, 10 misbehaving PVs are selected randomly for reputation updating in a period of time. The update period of PVs' reputation is 1 minute. The observation time of our simulation is 10 minutes. Initially, the misbehaving PVs pretend to act cooperatively. The probability of normally carrying out the consensus process reaches to 80\%. After 5 or 10 minutes, the misbehaving nodes show the misbehavior and the probability is dropped to 10\%. The reputation values under the LR scheme are higher than those under the SL scheme. By combining the reputation opinions and weight adequately by considering prior knowledge such as familiarity and timeliness, the SL scheme detects the misbehaviour more clearly. This means that our approach achieves a sensitive reputation updating which guarantees secure and efficient consensus in DBFT.

Fig. 5 shows the detection rate changes under different schemes over time. As time goes by, the proposed SL scheme can detect many more misbehaving PVs than the LR scheme. Moreover, the misbehaving PVs will be detected by the SL scheme more quickly compared with the LR scheme. When the threshold reputation value is 0.45, the detection rate under the SL scheme reaches 100\% faster than that with the LR scheme by 3 minute. With higher value of threshold, the misbehaving PVs will be easier to be detected. With higher and faster detection rate in the proposed SL scheme, the potential security threats can be detected and removed more effectively, which achieves a secure DBFT consensus process.

As shown in Fig. 6, we evaluate the correct probability of block verification with respect to different reputation thresholds. In the cases with a very low reputation threshold, the misbehaving PVs may launch the verification collusion attack. More than 1/3 misbehaving consensus nodes collude to generate false verification result for block generation. The correct probability of verified block means that the data block is correctly verified even in the presence of the verification collusion attack. When the reputation threshold is 0.45 and the base rate is 0.5, the correct probability under the proposed SL scheme is 36\% higher than that of the LR scheme. This indicates that the proposed SL can achieve a secure block verification, even when attackers launch internal active consensus nodes collusion.

\begin{figure}[t]\centering
  \includegraphics[width=0.5\textwidth,height=6.5cm]{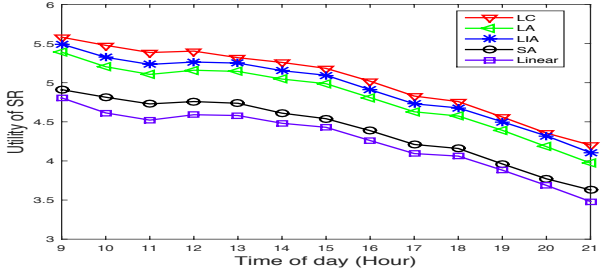}
  \caption{Utility of SR under different schemes versus time of day.}
  \label{a}
\end{figure}
\begin{figure}[t]\centering
  \includegraphics[width=0.5\textwidth,height=6.5cm]{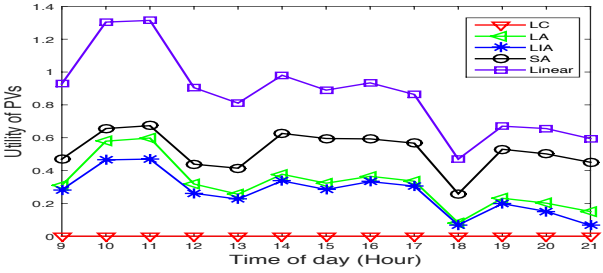}
  \caption{Utility of PVs under different schemes versus time of day.}
  \label{a}
\end{figure}

\subsection{Feasibility of Contract}
In order to verify the IR and the IC conditions of the proposed contract-based incentive scheme under asymmetric information scenario, the utilities of type-$3$, type-$5$ PVs with the proposed Lagrangian-based iterative scheme and LA scheme are shown in Fig. 7. Each type of PV selects all the contract $(\pi_j,f_j),j \in 1,2,...,N$, offered by the SR. Fig. 7 shows that each type of PV can achieve its maximum utility when it selects the contract that fits its own corresponding type, which indicates that the IC constraint is satisfied. For instance, the type-$3$ PVs achieve the maximum value only when it chooses the contract item $(\pi_{3},f_{3})$. If the type-$3$ PVs choose any other contract item $(\pi_j,f_j),j \in 1,2,...,N$, and $j \ne 3$, the utility of the type-$3$ PVs cannot reach the peak value compared with the contract $(\pi_{3},f_{3})$. Furthermore, it can be seen that each type of PVs selects the contract that fits its corresponding type and receives a nonnegative value, which suggests that the IR constraint is satisfied. Therefore, after choosing the best contract that is  designed for its own type, the type of the PVs will be revealed to the SR. In other s, by applying the proposed contract-based incentive scheme, the SR can be aware of the PVs' hidden information, such as its type, and thus overcome information asymmetry.

\subsection{Performance Evaluation With Different Time of a Day}
\begin{figure}[t]\centering
  \includegraphics[width=0.5\textwidth,height=6.5cm]{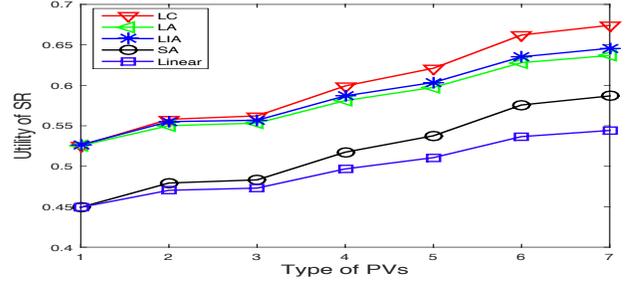}
  \caption{Utility of SR under different schemes for different type of PVs.}
  \label{a}
\end{figure}
\begin{figure}[t]\centering
  \includegraphics[width=0.5\textwidth,height=6.3cm]{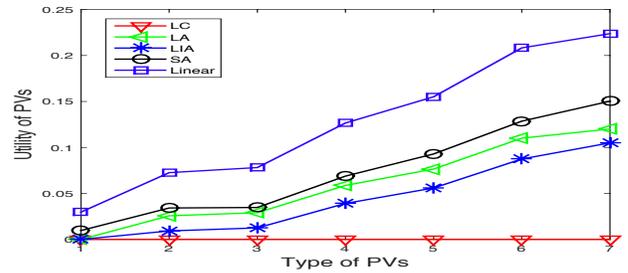}
  \caption{Utility of PVs under different schemes for different type of PVs.}
  \label{a}
\end{figure}
The parking behaviour varies significantly along the day and matches people's daily routines and habits. Fig. 8 compares the utility of SR with our proposed LIA scheme with other four schemes considering different time in a day. As shown in this figure, the utility of SR with all five schemes achieve peak value at $9$ A.M., with a second smaller peak occurring again around 12 A.M. This is because, most long-term parking happens in the early hours of the morning, which coincides with the hours at which most day jobs begin. The utility of SR achieves the highest value with the LC scheme, which can be seen as the upper bound. The proposed LIA scheme achieves the second best utility which can be seen as the upper bound under asymmetric information scenario. The proposed LIA scheme achieves better utility than the SA scheme. The reason is that, in the contract-based scheme, each contract is designed for the corresponding PV's type, and the utility of SR could be improved by binding the LDICs as described in \textbf{Lemma 5} and leave less surplus for the PVs. However, each type of PVs determines the selling price according to its own utility which tends to maximize its own utility in the Stackelberg game-based scheme. The linear pricing scheme provides the worst utility compared to the other four schemes.

Fig. 9 compares the utility of PVs under five schemes for different time in a day. As the figure shows, the utility of PVs grows rapidly at 10 A.M., reaches the peak values at 11 A.M and the minimum value at 6 P.M. This is because, most of the people begin their jobs in the early hours and return home in the afternoon, which leading to the long-term parking in the morning and the short-term parking in the afternoon. Furthermore, the PVs' utility with LC scheme remains zero all the time which is considered as the lower bound among five schemes. Since the PVs' type is observable to the SR under complete information scenario, the SR can extract the revenue from the PVs, which leaves the PVs zero utility. The PVs' utility with linear pricing scheme achieves the best utility compared with the other four schemes, followed by the SA scheme. The reason is similar to that for Fig. 8, each type of PVs has the freedom to maximize its utility function and thus can reserve more profit than contract-based schemes. The proposed LIA scheme gives the second worst performance in terms of PVs' utility. This is because, the PVs have limited contract to select from the SR and thus reserve less profit from the SR.

\subsection{Performance Evaluation With Different Types of PVs}
Fig. 10 illustrates system performance of SR's utility with respect to type of PVs. We randomly set a time in a day. Fig. 10 shows that it is profitable for the SR to motivate the higher type of PVs to contribute their underutilized computation resources to help offload computing tasks. That because the higher type of PVs are with higher probability to stay in the parking place during a short time period, which is more stable for computation offloading. Furthermore, Fig. 10 provides a clear gap between the contract-based schemes and those without contract-based schemes. This means, the contract-based scheme is more efficient than the other two schemes. Also, we can see that the contract-based scheme under complete information scenario brings the maximum utility for the SR, followed by the proposed LIA method-based iterative scheme. Under asymmetric information scenario, since the SR is not aware of the type of PVs, the designed IC-based contract can only bring a approximate optimal utility for the SR, which is upper bounded by that for the complete information scenario. The utility of SR under the Stackelberg game scheme outperforms the linear pricing scheme which gives the worst performance compared with all these five schemes. 

Fig. 11 shows system performance of PVs' utility with different types of PVs. As shown in Fig. 11, the utility of PVs always remains zero when the contracts are designed under complete information scenario. Except the contract-based scheme under complete information scenario, the other four schemes achieve higher utility of PVs with higher PVs' type. This is because, the higher type of PVs can provide more stable computation offloading services with the SR and thus gain more profit than the lower types. Similar to Fig. 8, the utility of PVs with linear pricing scheme gives the best performance among all these five schemes. This is because, each type of PVs can determine their performance and optimize their self-interest. The SA scheme comes the second one and outperforms the LIA schemes.

\section{Conclusion}
We have proposed a permissioned vehicular blockchain in VEC, called \emph{Parkingchain}, where the parked vehicles (PVs) can share their idle computational resources with service requesters (SRs). We have utilized the blockchain technology and designed a smart contract to achieve efficient resource sharing of PVs and secure service provisioning. A subjective logic-based DBFT consensus mechanism has been presented to enhance the consensus process in Parkingchain. For secure communications, we have leveraged the multi-weighted subject logic model to evaluate reputation values of PVs with higher accuracy. Further, we have proposed a contract-based incentive mechanism to model the decision process between a SR and PVs under asymmetric information scenario. Optimal contracts have been designed to reward the PVs with different parking behaviour and energy consumption for computation offloading while maximizing the SR's utility. Finally, simulation results have demonstrated that the proposed contract-based scheme is feasible and effective compared with traditional schemes. In our future work, in order to compute the reputation value more accurately, we will improve the DBFT consensus process by taking more factors and weights into consideration. Also, we will investigate on he application of machine learning for task allocation to containerized vehicles in Parkingchain.

\appendices
\section{Proof of Lemma 1}
We know that the types of PVs satisfy ${\theta _1} < {\theta _2} <  \cdot  \cdot  \cdot  < {\theta _j} <  \cdot  \cdot  \cdot  < {\theta _N}$. Combining IC constraints in (\ref{opt-simplified}a), we have
\begin{equation}
\begin{aligned}
{\theta _j}v\left( {{\pi _j}} \right) - e\kappa s \varepsilon {f_j}^2 & \ge {\theta _j}v\left( {{\pi _1}} \right) - e\kappa s \varepsilon {f_1}^2 \\
&\ge {\theta _1}v\left( {{\pi _1}} \right)- e\kappa s \varepsilon {f_1}^2 \ge 0.
\end{aligned}
\end{equation}
Therefore, if the IR constraint of type-$1$ PVs is satisfied, all PVs will satisfy the IR constraints. This completes the proof.

\section{Proof of Lemma 4}
According to IC constraints in (\ref{opt-simplified}a), we have the following two LDICs:
\begin{equation}
{\theta _{j+1}}v\left( {{\pi _{j+1}}} \right) - e\kappa s \varepsilon {f_{j+1}}^2  \ge {\theta _{j+1}}v\left( {{\pi _{j}}} \right) - e\kappa s \varepsilon{f_{j}}^2,
 \end{equation}
 \begin{equation}
{\theta _j}v\left( {{\pi _j}} \right) - e\kappa s \varepsilon {f_j}^2  \ge {\theta _{j}}v\left( {{\pi _{j-1}}} \right) - e\kappa s \varepsilon{f_{j-1}}^2.
 \end{equation}
 According to Lemma 2 and Lemma 3, if $\theta_j > \theta_k$, then $\pi_j > \pi_k$ and $f_j > f_k$, $\forall j>k, \in \left\{ {1,...,N} \right\}$. Based on (33), we have

\begin{equation}
\begin{aligned}
{\theta _{j + 1}}\left( {v\left( {{\pi _j}} \right) - v\left( {{\pi _{j-1}}} \right)} \right) & \ge {\theta _j}\left( {v\left( {{\pi _j}} \right) - v\left( {{\pi _{j-1}}} \right)} \right)\\
 & \ge e\kappa s \varepsilon {f_{j}}^2 - e\kappa s \varepsilon {f_{j-1}}^2.
\end{aligned}
\end{equation}

\noindent Combining (32) and (34), we have
\begin{equation}
\begin{aligned}
 {\theta _{j + 1}}v\left( {{\pi _{j+1}}} \right) -e\kappa s \varepsilon {f_{j+1}}^2& \ge {\theta _{j + 1}}v\left( {{\pi _j}} \right) - e\kappa s \varepsilon {f_{j}}^2 \\
 & \ge {\theta _{j + 1}}v\left( {{\pi _{j-1}}} \right)- e\kappa s \varepsilon{f_{j-1}}^2.
\end{aligned}
\end{equation}

\noindent Thus, we have
\begin{equation}
 {\theta _{j + 1}}v\left( {{\pi _{j+1}}} \right)- e\kappa s \varepsilon{f_{j+1}}^2  \ge {\theta _{j + 1}}v\left( {{\pi _{j-1}}} \right) - e\kappa s \varepsilon {f_{j-1}}^2.
\end{equation}

\noindent The inequality (36) could be extended to prove that all the DICs are hold. This proof can be shown as
\begin{equation}
\begin{aligned}
 {\theta _{j + 1}}v\left( {{\pi _{j+1}}} \right) - e\kappa s \varepsilon {f_{j+1}}^2 & \ge {\theta _{j + 1}}v\left( {{\pi _{j-1}}} \right) -e\kappa s \varepsilon{f_{j-1}}^2\\
 & \ge  \cdot  \cdot  \cdot  \\
 & \ge {\theta _{j+ 1}}v\left( {{\pi _{1}}} \right)- e\kappa s \varepsilon {f_{1}}^2.
\end{aligned}
\end{equation}

Since the proof of LUIC reduction is similar to that of reducing LDIC, we omit the prove here. Therefore, we conclude that the DICs and UICs hold with the monotonicity, the LDIC and the LUIC.

\section{Proof of Lemma 5}
First, we prove that the LDIC can be simplified as ${{\theta _j}v\left( {{\pi _j}} \right) - e\kappa s\varepsilon {f_j}^2 = {\theta _j}v\left( {{\pi _{j - 1}}} \right) - e\kappa s\varepsilon {f_{j - 1}}^2}$, which together with monotonicity can guarantee the LUIC hold. Suppose that the LDICs are satisfied by all type of PVs, and we have
\begin{equation}
\begin{aligned}
\begin{array}{*{20}{c}}
{{\theta _j}v\left( {{\pi _j}} \right) - e\kappa s\varepsilon {f_j}^2 \ge {\theta _j}v\left( {{\pi _{j - 1}}} \right) - e\kappa s\varepsilon {f_{j - 1}}^2}
\end{array},\\{\forall j \in \left\{ {2,...,N} \right\}}.
\end{aligned}
\end{equation}
Notice that the LDIC will still be guaranteed if both $\pi_j$ and $\pi_{j-1}$ are reduced by the same amount. With the aim to maximize it own utility, the SR will try it best to reduce the $\pi_j$ until the following equation
\begin{equation}
\begin{aligned}
\begin{array}{*{20}{c}}
{{\theta _j}v\left( {{\pi _j}} \right) - e\kappa s\varepsilon {f_j}^2 = {\theta _j}v\left( {{\pi _{j - 1}}} \right) - e\kappa s\varepsilon {f_{j - 1}}^2}
\end{array},\\{\forall j \in \left\{ {2,...,N} \right\}}.
\end{aligned}
\end{equation}
Note that the other types' LDIC will not be affected by this process.

Next, we prove that the LUIC will be hold if ${{\theta _j}v\left( {{\pi _j}} \right) - e\kappa s\varepsilon {f_j}^2 = {\theta _j}v\left( {{\pi _{j - 1}}} \right) - e\kappa s\varepsilon {f_{j - 1}}^2}$, ${\forall j \in \left\{ {2,...,N} \right\}}$ and the monotonicity hold. Since we have ${{\theta _j}v\left( {{\pi _j}} \right) - e\kappa s\varepsilon {f_j}^2 = {\theta _j}v\left( {{\pi _{j - 1}}} \right) - e\kappa s\varepsilon {f_{j - 1}}^2}$, ${\forall j \in \left\{ {2,...,N} \right\}}$, we have
\begin{equation}
e\kappa s\varepsilon \left( {{f_j}^2 - {f_{j-1}}^2} \right) = {\theta _j}\left( {v\left( {{\pi _{j-1}}} \right) - v\left( {{\pi _{j-1}}} \right)} \right).
\end{equation}

Due to the monotonicity, if $\theta_j \ge \theta_{j-1}$, then $\pi_j \ge \pi_{j-1}$, we have

\begin{equation}
{\theta _j}\left( {v\left( {{\pi _j}} \right) - v\left( {{\pi _{j - 1}}} \right)} \right) \ge {\theta _{j - 1}}\left( {v\left( {{\pi _j}} \right) - v\left( {{\pi _{j - 1}}} \right)} \right).
\end{equation}
Combining (40) and (41), we have
\begin{equation}
\begin{aligned}
{\theta _j}\left( {v\left( {{\pi _j}} \right) - v\left( {{\pi _{j - 1}}} \right)} \right) &= e\kappa s\varepsilon \left( {{f_j}^2 - {f_{j - 1}}^2} \right) \\ &\ge {\theta _{j - 1}}\left( {v\left( {{\pi _j}} \right) - v\left( {{\pi _{j - 1}}} \right)} \right),
\end{aligned}
\end{equation}
and thus we have
\begin{equation}
v\left( {{\pi _{j - 1}}} \right) - e\kappa s\varepsilon {f_{j - 1}}^2 \ge {\theta _{j - 1}}v\left( {{\pi _j}} \right) - e\kappa s\varepsilon {f_j}^2,
\end{equation}
which is exactly the LUIC constraint. Therefore, the LUIC can be removed from the constraints in (\ref{opt-simplified}).

\section{Proof of Lemma 6}
To obtain the optimal contracts under complete information scenario, we replace $f_j$ in $U_{SR_j}$ with (23) as follows:
\begin{equation}
{U_{S{R_j}}} = {\beta _j}{\theta _j}\left[ {\rho \left( {\frac{{\kappa s}}{{{f_{local}}}} - \frac{{\kappa s}}{{\sqrt {\frac{{{\theta _j}v\left( {{\pi _j}} \right)}}{{e\kappa s\varepsilon }}} ,}} - \frac{s}{{{r_j}}}} \right) - {\pi _j}} \right].
\end{equation}
Then, we first take a derivative of (44) as follows:
\begin{equation}
\frac{{d{U_{S{R_j}}}}}{{d{\pi _j}}} = \frac{{{\beta _j}\theta _j^2\rho }}{{2e\varepsilon }}{\left( {\frac{{{\theta _j}v\left( {{\pi _j}} \right)}}{{e\kappa s\varepsilon }}} \right)^{ - \frac{3}{2}}}\frac{{dv\left( {{\pi _j}} \right)}}{{d{\pi _j}}} - 1.
\end{equation}
Since $v\left(  \cdot  \right)$ is a strictly increasing function of $\pi_j$, where $v\left( 0 \right) = 0,v'\left( {{\pi _j}} \right) > 0$, and $v''\left( {{\pi _j}} \right) < 0, j \in \{1,...,N\}$. Further, we observe the derivative of (45) and have
\begin{equation}
\begin{array}{l}
\frac{{{d^2}{U_{S{R_j}}}}}{{d\pi _j^2}} =  - \frac{{3{\beta _j}\theta _j^3\rho }}{{4{e^2}{\varepsilon ^2}\kappa s}}{\left( {\frac{{{\theta _j}v\left( {{\pi _j}} \right)}}{{e\kappa s\varepsilon }}} \right)^{ - \frac{5}{2}}}{\left( {\frac{{dv\left( {{\pi _j}} \right)}}{{d{\pi _j}}}} \right)^2} + \\
\frac{{{\beta _j}\theta _j^2\rho }}{{2e\varepsilon }}{\left( {\frac{{{\theta _j}v\left( {{\pi _j}} \right)}}{{e\kappa s\varepsilon }}} \right)^{ - \frac{3}{2}}}\frac{{{d^2}v\left( {{\pi _j}} \right)}}{{d\pi _j^2}} < 0,
\end{array}
\end{equation}
which indicates that $U_{SR_j}$ is a concave function on $\pi_j$, and there exists the ${{\bar \pi }_j}$ that is optimal and maximizes $U_{SR_j}$ with given ${{\bar \pi }_{j}}$.

\section{Proof of Lemma 7}
To obtain the locally optimal contracts, we first consider $U_{SR_1}$ with type-$1$ PVs as follows:
\begin{equation}
{U_{S{R_1}}} = {\beta _1}{\theta _1}\left[ {\rho \left( {\frac{\kappa s}{{{f_{local}}}} - \frac{\kappa s}{{\mu_p{f_1}}} - \frac{s}{{{r_1}}}} \right) - {\pi _1}} \right].
\end{equation}

We replace $f_1$ with $\pi_1$, according to $f_1={\frac{1}{{{\mu _p}}}\sqrt {\frac{{{\theta _1}v\left( {{\pi _1}} \right)}}{{e\kappa s\varepsilon }}} }$. Then, we can obtain the locally optimal contract $\left\{ {{{\dot f}_1},{{\dot \pi }_1}} \right\}$ for type-$1$ PVs by solving ${{d{U_{S{R_1}}}} \mathord{\left/{\vphantom {{d{U_{S{R_1}}}} {d{\pi _1}}}} \right.\kern-\nulldelimiterspace} {d{\pi _1}}} = 0$.

Further, we consider the locally optimal contracts when $2 \le j \le N$. We replace $f_j$ with $\pi_j$ in $U_{SR_j}$ with (24). Then, we first take a derivative of $U_{SR_j}$ as follows:
\begin{equation}
{\frac{{d{U_{S{R_j}}}}}{{d{\pi _j}}} = \frac{{{\beta _j}\theta _j^2\rho }}{{2e\varepsilon }}{{\left( {\frac{{{\theta _1}v\left( {{\pi _1}} \right) + \sum\nolimits_{k = 2}^j {{\Delta _k}} }}{{e\kappa s\varepsilon }}} \right)}^{ - \frac{3}{2}}}\frac{{dv\left( {{\pi _j}} \right)}}{{d{\pi _j}}} - 1},
\end{equation}
where ${\Delta _k} = {\theta _k}\left[ {v\left( {{\pi _k}} \right) - v\left( {{\pi _{k - 1}}} \right)} \right]$. Since $v\left(  \cdot  \right)$ is a strictly increasing function of $\pi_j$, where $v\left( 0 \right) = 0,v'\left( {{\pi _j}} \right) > 0$, and $v''\left( {{\pi _j}} \right) < 0, j \in \{1,...,N\}$. We observe the derivative of (48) and have
\begin{equation}
\begin{array}{*{20}{l}}
{\frac{{{d^2}{U_{S{R_j}}}}}{{d\pi _j^2}} =  - \frac{{3{\beta _j}\theta _j^3\rho }}{{4{e^2}\kappa s{\varepsilon ^2}}}{{\left( {\frac{{{\theta _1}v\left( {{\pi _1}} \right) + \sum\nolimits_{k = 2}^j {{\Delta _k}} }}{{e\kappa s\varepsilon }}} \right)}^{ - \frac{5}{2}}}{{\left( {\frac{{dv\left( {{\pi _j}} \right)}}{{d{\pi _j}}}} \right)}^2}}\\
{ + \frac{{{\beta _j}\theta _j^2\rho }}{{2e\varepsilon }}{{\left\{ {\frac{{{\theta _1}v\left( {{\pi _1}} \right) + \sum\nolimits_{k = 2}^j {{\Delta _k}} }}{{e\kappa s\varepsilon }}} \right\}}^{ - \frac{3}{2}}}\frac{{{d^2}v\left( {{\pi _j}} \right)}}{{d\pi _j^2}} < 0},
\end{array}
\end{equation}
which indicates that $U_{SR_j}$ is a concave function on $\pi_j$, and there exists the ${{\dot \pi }_j}$ that is locally optimal and maximizes $U_{SR_j}$ with given ${{\dot \pi }_{j-1}}$. For $j \ge 2$, ${{\dot \pi }_j}$ can be obtained by solving ${{d{U_{S{R_j}}}} \mathord{\left/{\vphantom {{d{U_{S{R_j}}}} {d{\pi _j}}}} \right.\kern-\nulldelimiterspace} {d{\pi _j}}} = 0$ and $\dot f_j$ can be solved according to (24).

\bibliography{myreference}

\end{document}